\documentstyle[aps,prl]{revtex}
\begin{document}
\large

\title{Direct hadron production and radial flow reduction as quark-gluon
plasma signature.}

\author{D. Yu. Peressounko}

\address{Russian Research Center
"Kurchatov Institute", Kurchatov sq. 1, Moscow, Russia}

\maketitle

\begin{abstract}
It is shown that due to contribution of the hadrons, emitted directly from
the inner hot region of central heavy ion collision, radial flow of the
final state hadrons is rather sensitive to the creation of the quark-gluon
plasma with mixed phase.
\end{abstract}

\pacs1{25.75.-q}

One of the main issues of the physics of ultrarelativistic heavy ion
collisions is creation and registration of new state of matter --
quark-gluon plasma (QGP). Several signatures of QGP creation in these
collisions have been proposed so far: strangeness enhancement, charm
suppression, direct photons and dileptons emission, etc. (see e.g.
contributions in \cite{QGPsignatures}). However, detail considerations show,
that nearly all proposed signals are not unique features of QGP formation,
instead, they can be simulated more or less successfully by hot and dense
hadronic matter. So that, the problem of identification of QGP in the
heavy ion collisions is still open now.

In this letter we discuss radial flow reduction in central heavy ion
collisions as a possible signal of QGP formation. It was pointed out several
times (see e.g. \cite{Hydro}), that hadronic matter going through phase
transition expands much slowly than pure hadronic gas. However, to compare
flows, created in the collisions with and without phase transition it is
necessary to specify initial conditions for hydrodynamic expansion. They can
not be evaluated exactly, and have to be chosen from some speculations.
Usually, one uses initial conditions identical both for evolution with phase
transition and without it, and finds, that collective flow, created in the
first case of evolution is lower than in the second. Within this approach one
can not simultaneously reproduce experimental hadronic distributions both in
evolution with and without phase transition, so that one has to assume {\it
ab initio} wether or not phase transition takes place in collision. Below we
accept another, more consistent to our opinion approach, in which one does
not make beforehand assumption about phase transition, and demands that
distributions of final state hadrons reproduce ones, measured on experiment
in both cases of evolution. Within this approach, in the frame of pure
hydrodynamics one obtains equal collective flows in both cases, but, as we
show below, if one takes into account emission of hadrons from central hot
part of the collision, then one finds that collective flow decreases if phase
transition takes place in the collision.

In the paper \cite{We} we have shown, that due to the comparability of the
size of the region, occupied by hot and dense matter created in the heavy ion
collision, and free path length of a hadron in it, final state hadrons can
originate not only from thin freeze-out hypersurface due to hadronic gas
decay (thus originated hadrons we call {\it freeze-out} hadrons below), but
also due to emission from the whole volume of dense matter and partial
escaping from the hot region without rescattering (this kind of final state
hadrons we call {\it direct} hadrons below). These two kinds of final state
hadrons test different stages of the collision: direct hadrons deliver
information about central hot part of the collision, while freeze-out hadrons
-- about freeze-out stage of collision. In a usually used pure hydrodynamic
model one neglects direct hadron emission and considers only freeze-out
hadrons, thus testing only stage of decaying of the hadronic gas.

Having in mind possibility to extract information about inner part of
the collision by use of the direct hadrons, let us compare evolutions of the
collision with and without QGP creation. When dealing with hydrodynamic
expansion of the system, which undergoes first order phase transition, one
finds, that while matter is in the mixed phase, its collective velocity does
not increase, because the pressure in the mixed phase is independent on local
energy density. In contrast to this, in the system without first order phase
transition pressure gradient is always present and collective velocity always
increase. This difference in collective velocity distributions in expanding
matter reveal itself in the collective velocity distributions of the direct
hadrons: {\it direct hadrons, emitted in the collision with phase transition,
have lower collective velocities than ones, emitted in the collision without 
phase transition}. In other words, if mixed phase is created in the 
collision, then it works like some kind of thermal bath, which abundantly 
emits direct hadrons with constant low collective velocity and high 
temperature.

To evaluate collective velocity distributions of final state hadrons in the
central heavy ion collision, we use the following model. The hot matter,
created during penetration of colliding nuclei through each other,
hydrodynamicaly expands, emitting direct hadrons from the whole volume,
occupied by it: from QGP -- due to emission of quarks and gluons and their
hadronization on its surface, from mixed phase and from surrounding hadronic
gas before beginning of freeze-out. We take into account possible
rescattering of flying out direct hadrons using probability to escape:
$P=\exp \left\{ -\int \lambda ^{-1}(x)\,dx\right\} $, where integration is
performed along the straight line -- path of the particle in the hot matter,
and $\lambda (x)$ - depending on local energy density free path length. To
evaluate hydrodynamic expansion and to calculate energy density distribution,
which is necessary for free path length calculations, we use scaling
(Bjorken) hydrodynamics with transverse expansion modified to take into
account energy loss due to direct hadron emission. As far as we consider
radial expansion in central collisions, it is sufficient to use 1+1
dimensional hydrodynamics. Hadronization of the quark or gluon flied out
from QGP or mixed phase we describe as decay of the string, pulled out by
this quark or gluon, onto one hadron with the only restriction: energy of the
final hadron must be lower than energy of the initial quark or gluon. In this
letter we do not consider emission of heavy hadrons and resonances, but
restrict ourselves to the emission of pions. Also in equation of state used
in hydrodynamic expansion we suppose that hadronic gas consists only of
pions. Calculating collective velocity distributions of hadrons, we assume
that emitted hadron has the same collective velocity as the elementary
volume, from which it is emitted. In this letter we consider only transverse
collective velocity, but it is clear, how to apply this model to the
longitudinal one. Solving hydrodynamic equations numerically, we use various
equations of state (EoS), depending on wether phase transition takes place in
the collision. If we assume that QGP is not created, then we use EoS of ideal
gas -- short-dashed line on Fig.\ \ref{fig1}. Otherwise we use EoS with first
order phase transition (solid line on Fig.\ \ref{fig1}). Fluctuations in the
final size system can be simulated by 'smoothing' of the EoS. This effect
could be important in our calculations, so we test sensitivity of our
predictions to the sharpness of EoS -- we also use third kind of EoS with
smoothed first order phase transition (long-dashed line on Fig.\ \ref{fig1}).
More detailed description of the calculation of direct hadron emission can be
find in \cite{WeModel}.

Within this model we have the following set of free parameters, which values
have to be chosen from some considerations: $T_{in}$ -- initial temperature,
$R_{in}$ -- initial radius of Bjorken cylinder, $\tau _{in}$ -- initial time,
$T_c$ -- transition temperature and $T_{freez}$ -- freeze-out temperature.

We apply our model to the recently experimentally studied central Pb+Pb
collisions at $158\;A\cdot GeV$ -- i.e. we adjust parameters of the model to
reproduce measured $p_t$ distribution and multiplicity of pions at
midrapidity ($y\approx 3$). So we have two conditions:  multiplicity and
slope of the $p_t$ distribution, and five (four in the case of pure hadronic
gas) free parameters, therefore, we can vary values of parameters in the
reasonable limits still reproducing multiplicity and $p_t$ distribution.  In
the case of QGP creation, both for `sharp' and `smooth' EoS we take
$T_{in}=180\;MeV, \;R_{in}=6.5\;fm,\;\tau _{in}=1.3\;fm/c,\; T_c=150\; MeV,\;
T_{freez}=140\; MeV$ , and for pure hadronic gas evolution -- $T_{in}=180\;
MeV,\; R_{in}=12\;fm,\; \tau _{in}=6.5\; fm/c,\;T_{freez}=140\;MeV$.

Resulting transverse collective velocity distributions of final state hadrons
for four possible modes of evolution are shown on Fig.\ \ref{fig2}:
Evolution without QGP creation (dash-dotted line), with QGP creation and
`sharp' EoS (solid line), with QGP creation and `smooth' EoS (dashed line)
and with QGP creation but without direct hadron emission (dotted line).
Transverse collective velocity distributions, obtained with assumption of
pure hadronic gas evolution and evolution with QGP creation without direct
hadron emission are very similar: in both cases main contributions comes from
region $\beta _r\sim 0.6$, although in the case of pure hadronic gas an
additional bump at low $\beta _r$ appears due to direct hadron contribution.
So, if we do not take into account direct hadron emission, then evolution
with and without QGP formation indeed can not be distinguished.  Otherwise,
if we consider direct hadron emission, then transverse collective velocity
distribution, calculated for EoS without phase transition spectacularly
differ from ones, calculated for EoS both with `sharp' and `smooth' phase
transition. In the last two cases distributions are quite similar:  both for
evaluation with `sharp' and `smooth' EoS the narrow peak at $\beta _r \sim
0.1$ appears due to direct hadron contribution and wide bump at $ \beta _r
\sim 0.6$ -- due to freeze-out hadrons contribution. In these evaluations the
system is assumed initially in the pure QGP phase, so that mixed phase
possess small collective velocity during expansion of pure QGP, and peak from
direct hadron contribution is shifted with respect to zero. As we have
already noted, we have some freedom in choosing of the particular values of
parameters, hence, shown distributions can not be considered as a strict
predictions, but rather as typical for four possible cases of evolution. For
instance, increasing of the difference $T_{in}-T_c$ leads to increasing of
transverse collective velocities of direct hadrons. Increasing of the
difference $T_c-T_{freez}$ results in increasing of ones of freeze-out
hadrons. However, in view of restriction on resulting $p_t$ distribution, we
can not shift contributions of direct and freeze-out hadrons too far from the
shown positions: e.g. for any reasonable values of parameters, peak of direct
hadrons contribution appears below $\beta _r\sim 0.2$.

Transverse collective velocity distribution of the final state hadrons in
the form, shown on Fig.\ \ref{fig2} can not be directly measured at experiment.
Nevertheless, presence or absence of the contribution of the hadrons with low
transverse collective velocity results in variation of the average transverse
collective velocity ($<\beta _r>$), which can be extracted from experimental
data. We test the sensitivity of the average transverse collective velocity
to the `amount' of QGP, created in the collision, i.e. to the initial energy
density. We fix $T_c=180\;MeV$, $T_{freez}=140\;MeV$ and vary initial energy
density, adjusting initial volume to reproduce multiplicity of pions at
midrapidity and $p_t$ distribution. Results of these calculations are shown on
Fig.\ \ref{fig3}. We find that $<\beta _r>$ decrease with increasing of the
initial energy density until $E_{in}\sim 2.2\;GeV/fm^3$, when it begins to
increase. That is, while the space-time volume, occupied by mixed phase
increase, the number of direct hadrons emitted from it increase too, reducing
$<\beta _r>$. As the energy density becomes higher than $2.2\;GeV/fm^3$,
pure QGP is created in the center of the collision and accelerates the
surrounding mixed phase, increasing $<\beta _r>$.

Within our model we can not evaluate dependence of the initial energy
density on the beam energy. Nevertheless, assuming initial
energy density to be monotonous function of beam energy, we can made a
qualitative prediction: If one consider the average transverse collective
velocity as a function of beam energy, then in the case of EoS with phase
transition, the specific minimum appears, while in the case of EoS without
it we expect, that $<\beta _r>$ monotonously increases with saturation. The
decreasing in the case of phase transition begins when mixed phase is created
in the collision, while the minimum is reached when pure QGP is created.

Extraction of average transverse collective velocity from experimental data
was already done by several authors for Ni+Ni, Au+Au, Si+Au, S+Au and Pb+Pb
collisions at beam energies $10^{-1}-10^2A\cdot GeV$. These results were
summarized by Herrmann \cite{Herrmann} and Xu \cite{Xu}. Both authors
conclude, that $<\beta _r>$ increase up to AGS beam energy ($\sim 10A\cdot
GeV$), while there is no agreement in the analysis of the present SpS data
(beam energy $\sim 10^2 A\cdot GeV$). Herrmann reported that average
transverse collective velocity decrease from $ <\beta _r>\approx 0.44$ at AGS
energy to $<\beta _r>\approx 0.27$ at SpS energy, while Xu concluded that
this is just a saturation at AGS energy at the level $<\beta _r>\approx 0.4$.
In this letter we can not make quantitative comparison with these data, but
mention that we obtain reasonable values of $<\beta _r>$, comparable with
those, discussed by Herrmann and Xu.

To summarize, we consider new possible signal of QGP creation in the heavy
ion collision. Due to comparability of the sizes of the hot matter and free
path length of the hadron in it, direct hadrons, emitted by QGP and by mixed
phase, can partially escape from the hot region and deliver information about
central part of the collision. We use this ability to test the presence of
the mixed phase in the collision: due to the mixed phase does not accelerate,
direct hadrons emitted from it have essentially lower collective velocity,
than in the collision without mixed phase creation, and thus average
collective velocity decreases. We calculate transverse collective velocity
distributions of final state hadrons in the central Pb+Pb collision at 158
$A\cdot GeV$ for various EoS, and demonstrate, that direct hadrons contribute
mainly into region $\beta _r \sim 0.1$, while freeze-out hadrons contribute
to the region $\beta _r\sim 0.6 $. We evaluate dependence of the average
transverse collective velocity on the initial energy density.  In the case of
phase transition it has specific minimum, while in the case of absence of
phase transition we expect monotonous increasing.  Available preliminary
experimental results are treated by different authors both as decreasing or
as just saturating of $<\beta _r>$, so this problem is open now.

I pleased to thanks Yu. E. Pokrovsky for inspiring discussions and
continuous support.

\begin{figure}
\caption{Equations of state used in evaluations: without phase transition
(dotted line), with 'sharp' phase transition (solid line) and with
'smooth' phase transition (dashed line).}
\label{fig1}
\end{figure}

\begin{figure}
\caption{Transverse collective velocity distributions of final state hadrons
for evaluations without QGP creation (dash-dotted line), with QGP creation
and `sharp' EoS (solid line), with QGP creation and `smooth' EoS (dashed
line) and with QGP creation but without direct hadron emission (dotted line).
}
\label{fig2}
\end{figure}

\begin{figure}
\caption{Predicted dependence of the average transverse collective velocity of
hadrons on initial energy density in the case of phase transition.}
\label{fig3}
\end{figure}
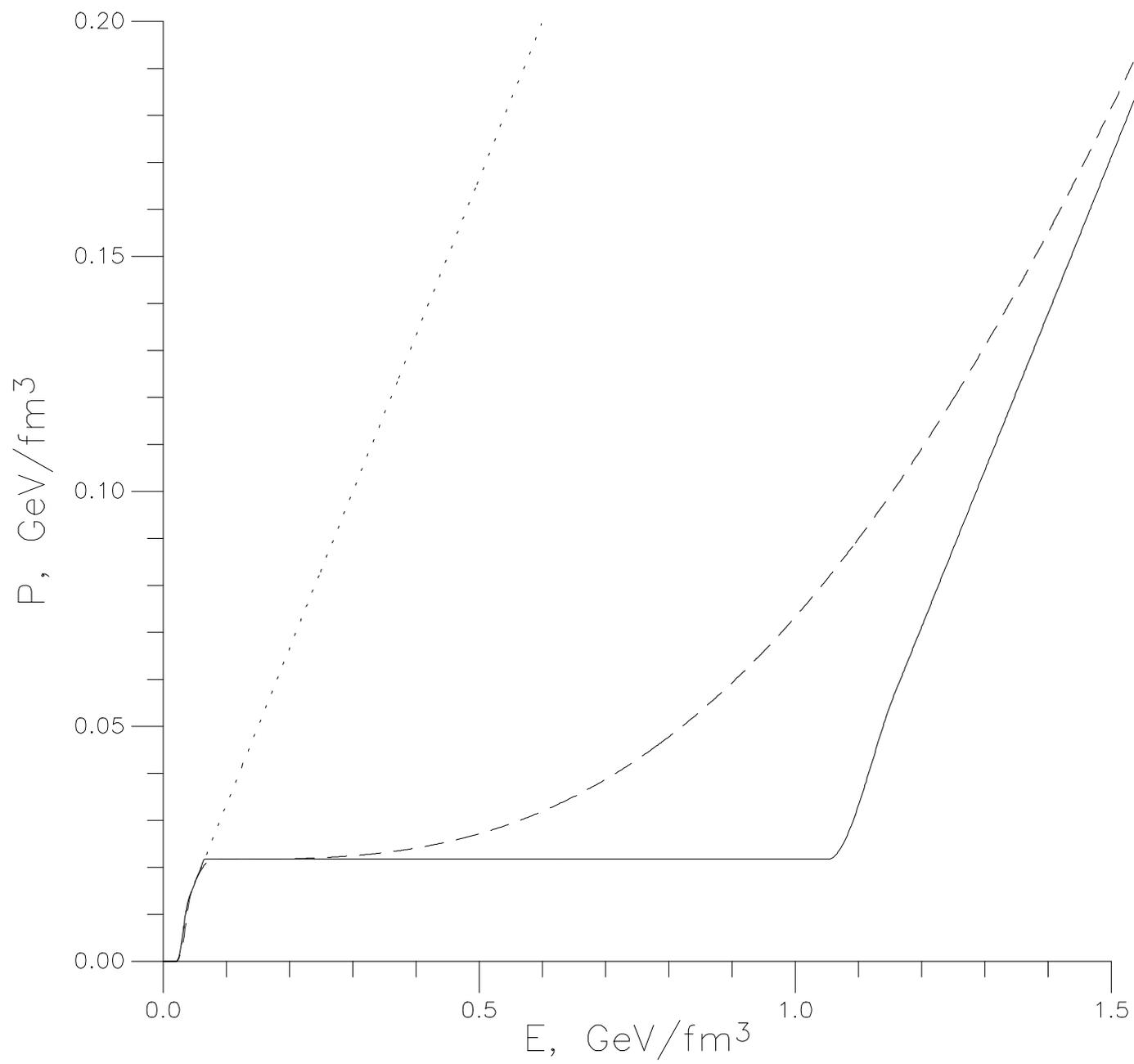
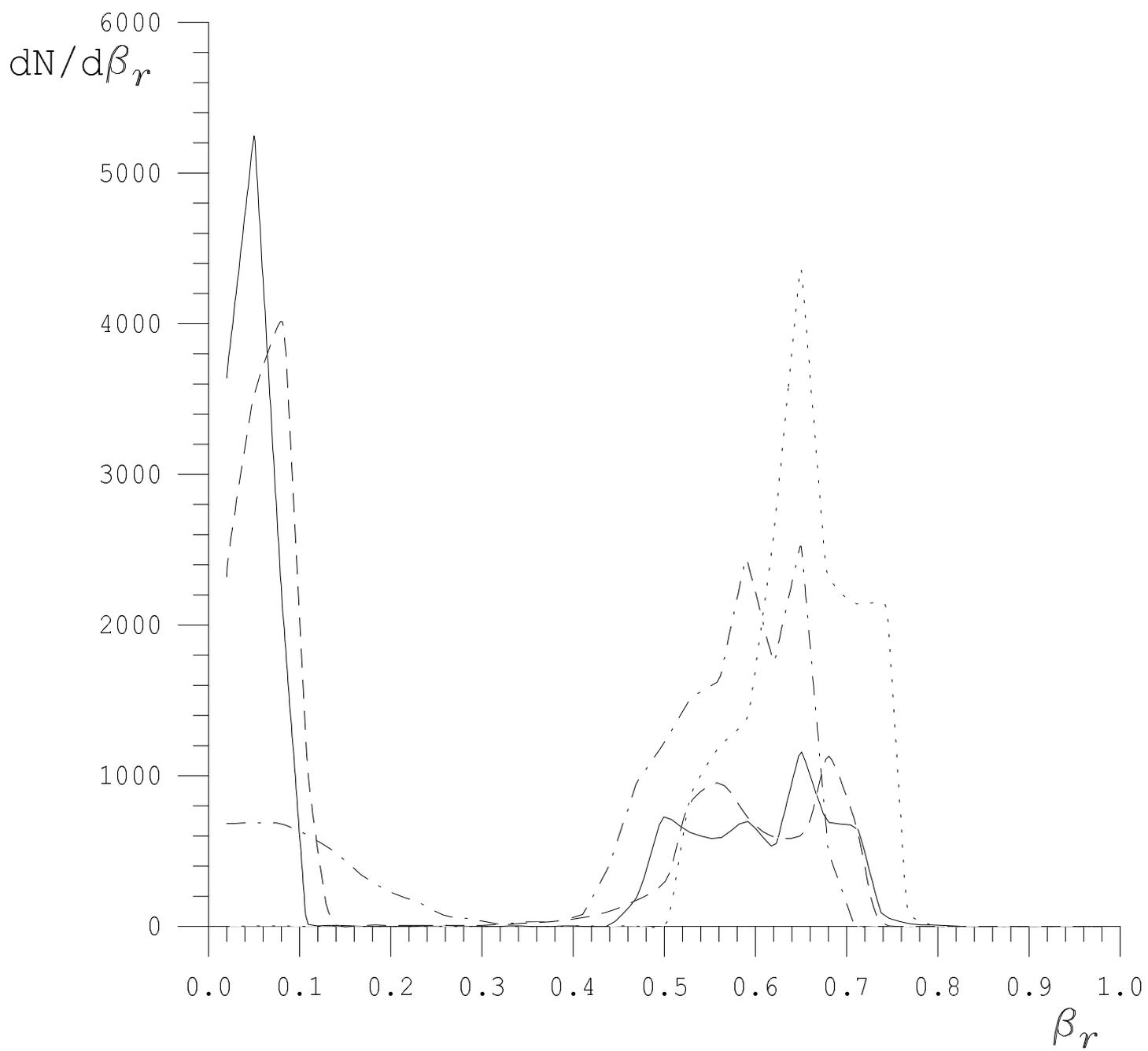
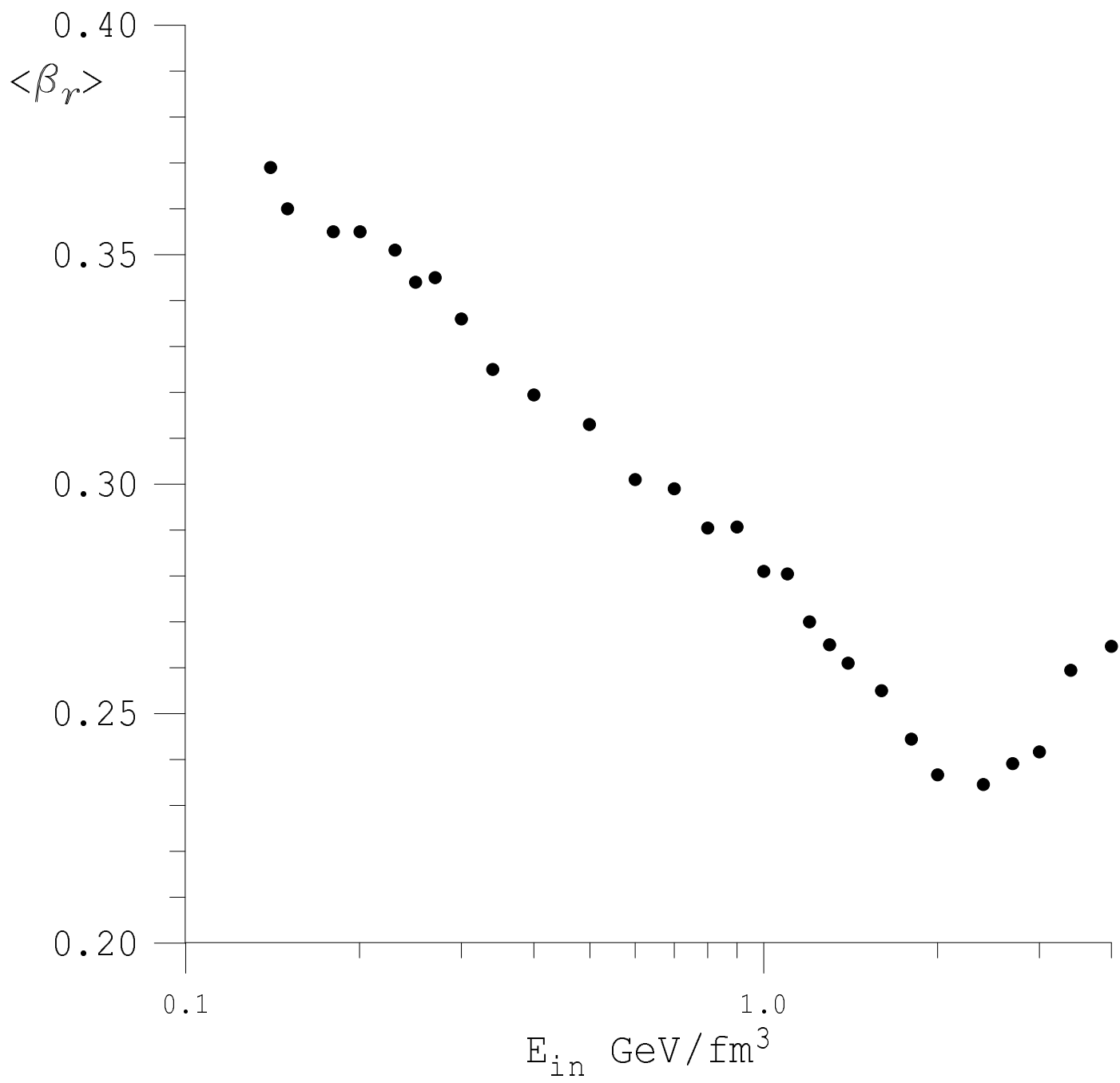
\end{document}